\documentclass[conference]{IEEEtran}
\usepackage{amsmath,graphicx}
\usepackage[affil-it]{authblk}
\usepackage{cite}
\usepackage{booktabs}
\usepackage{color}
\usepackage{url}
\usepackage{placeins}
\usepackage{stfloats}

\title{Assessing the Potential Impact of Direction-Dependent HRTF Selection on Sound Localization Accuracy}

\author[1]{Sapir Goldring}
\author[2]{Zamir Ben Hur}
\author[2]{David Lou Alon}
\author[1]{Boaz Rafaely}

\affil[1]{School of Electrical and Computer Engineering, Ben Gurion University of the Negev}
\affil[2]{Reality Labs Research, Meta, Redmond, WA 98052, USA}
\begin{document}

\maketitle
\begin{abstract}
This study investigates the approach of direction-dependent selection of Head-Related Transfer Functions (HRTFs) and its impact on sound localization accuracy. For applications such as virtual reality (VR) and teleconferencing, obtaining individualized HRTFs can be beneficial yet challenging, the objective of this work is therefore to assess whether incorporating HRTFs in a direction-dependent manner could improve localization precision without the need to obtain individualized HRTFs. A localization experiment conducted with a VR headset assessed localization errors, comparing an overall best HRTF from a set, against selecting the best HRTF based on average performance in each direction. The results demonstrate a substantial improvement in elevation localization error with the method motivated by direction-dependent HRTF selection, while revealing insignificant differences in azimuth errors.

\end{abstract}

\section{Introduction}
\label{sec:intro}
The Head Related Transfer Function (HRTF) represents the acoustic response between a distant sound source and the sound pressure at a listener's ear \cite{rafaely2010interaural,xie2013head}. 
Since individuals possess a unique anatomical ear structure and dimensions, the HRTF varies between individuals, in addition to changing along direction  \cite{majdak2007multiple}.
For applications such as Virtual Reality (VR), teleconferencing, and hearing aids \cite{madmoni2018direction}, where accurate sound spatialization is critical, the rendering of the auditory scene relies on these personalized HRTFs \cite{poirier2020assessing}. However, obtaining individual HRTFs is often impractical due to the time-intensive measurement process, leading to the utilization of non-individual HRTFs from existing databases \cite{katz2012perceptually}.

The selection of suitable HRTFs from a database typically involves conducting listening tests where the most appropriate HRTF is assigned to each individual based on perceptual criteria \cite{iwaya2006individualization,andreopoulou2016subjective,katz2012perceptually}. Various approaches are employed, ranging from participants rating sound trajectory relative to provided descriptions \cite{katz2012perceptually}\cite{andreopoulou2016subjective}, to tournament-style selection methods \cite{iwaya2006individualization}. Some works advocate for concentrating exclusively on source pathways and positions within the horizontal plane, where binaural cues such as the inter-aural time difference (ITD) and inter-aural level difference (ILD) hold prominence \cite{seeber2003subjective}. Since they vary less between listeners, effective azimuth localization can often be achieved even when using generic HRTFs  \cite{romigh2014you}. This highlights the need for a selection that is based on a perceptual test that considers a range of directions of arrival, including both elevation and azimuth.
Moreover, current HRTF selection methods aim to find the ``single most suitable" HRTF. However, HRTFs are frequency and direction dependent functions, thus a single HRTF may not provide the best fit in all directions and frequencies.
 
In this study, our objective is to assess the validity of the current HRTF selection approaches by comparing sound localization accuracy when choosing a single HRTF filter for a listener with an alternative approach of selecting the best-fitting HRTF for each specific Direction-of-Arrival (DOA), both in elevation and azimuth. The structure of this paper unfolds as follows: First, we introduce a database reduction method, grounded in the observation of spectral notches within HRTFs, typically appearing above 3 kHz \cite{poirier2020assessing}\cite{iida2014personalization}. Next, we outline the design of a listening experiment. Finally, we present our evaluation methods and the obtained results. 
The results of this study show that selecting different HRTFs for different directions, may provide improved individualization in terms of sound localization.

\section{Methods}

To detect the impact of direction dependent HRTF selection on sound localization accuracy, a localization listening test was conducted. This section, structured into three parts, outlines the methodology employed throughout the test.

The first part focuses on the description and reduction of the HRTF database. This includes an overview of the database used for the experiment and a detailed account of the reduction method applied to select a manageable number of HRTFs suitable for the localization experiment. 

The subsequent sections will delve into the specifics of the experiment setup and the three distinct methods for HRTF selection examined in this paper.

\subsection{HRTF dataset}
\label{sec:format}

The study relies on an internal HRTF dataset, which includes HRTFs from 96 individuals measured across 612 directions\cite{cuevas2019evaluation}. Notably, some HRTFs were measured multiple times (3 to 4 repetitions).

To streamline the sound localization experiment, the objective was to reduce the HRTF database to just five selected HRTFs. The primary goal was to ensure perceptual diversity among these chosen HRTFs, an approach aligned with the concept of ``perceptually orthogonal'' representations \cite{katz2012perceptually}. Given the limitations of traditional objective measures \cite{wightman1993multidimensional}, an alternative approach was selected, focusing on the first spectral notch within the HRTF.

Spectral notches exhibit variations among individuals, resulting in distinct perceptions of sound elevation \cite{xie2013head}. This suggests that when aiming to ensure perceptual differentiation among selected HRTFs concerning sound localization, it may be advisable to opt for HRTFs with distinct notches, positioned at a noticeable frequency gap from each other.  The first spectral notch of all HRTFs within the database was determined using the methodology outlined in \cite{iida2014personalization}. Subsequently, the mean and standard deviation among repetitions were computed, resulting in the selection of five HRTFs for this experiment. These selected HRTFs exhibited both low standard deviation, ensuring consistency, and a diverse range of mean values in the first notch frequency. Table \ref{tab:hrtf-stats} provides a description of both the mean and standard deviation (std) values for the selected HRTFs, alongside their corresponding subject IDs.

\begin{table}[t]
\caption{Selected HRTFs' Notch Frequencies and Standard Deviations.}
\label{tab:hrtf-stats}
\begin{center}
\begin{tabular}{ccc}
\toprule
ID & Notch Mean Frequency [kHz] & Notch Std [Hz]  \\
\midrule
5  & 9.4118 & 0\\
10 & 8.4706 & 266.2049\\
17 & 8.1568 & 54.3388 \\
42 & 7.5764 & 66.5512\\
63 & 8.0940 & 133.1025\\
\bottomrule
\end{tabular}
\label{tb:firsttb}
\end{center}
\end{table}

\subsection{Localization Experiment setup}

\label{sec:pagestyle}
The localization experiment was conducted using a Virtual Reality (VR) headset, with specific details outlined as follows \cite{warnecke2022hrtf}. Participants utilized the Meta Quest 2 headset for visualizations and wore AKG K702 headphones for sound immersion.

During the experiment, participants were presented with a black screen featuring sphere grids to aid in their spatial orientation. This visual context allowed participants to gain a sense of the direction they were pointing.

Auditory stimuli were delivered in the form of 3-second far-field unfiltered white noise, which was simulated from various DOAs and played in a free field environment.

The reference point for head orientation was established as 0 degrees azimuth and 0 degrees elevation, indicating the front-facing direction. In this system, elevation spans from -90 degrees (down) to 90 degrees (up), while azimuth ranges from 0 degrees (front) to 360 degrees (full circle).

For the investigation, we specifically targeted five distinct elevations: -47, -27, 0, 27, and 65 degrees. These elevations were symmetrically positioned at azimuths of either 40 degrees (left) or 320 degrees (right), representing the left and right directions, respectively. Azimuth values were randomized throughout the experiment.

Participants were exposed to the white noise and instructed to orient their heads in the direction from which they perceived the sound. The Quest 2 built-in head tracker was leveraged to precisely capture each participant's head pointing direction. To enhance data reliability, each combination of DOA and HRTF was repeated four times, resulting in a total of 100 experiment repetitions.
Besides monitoring the subjects' pointing direction, the head tracker introduced dynamic elements to the experiment. Participants were able to move their heads freely. If they exceeded a 15-degree deviation from the reference point in any direction, the stimuli would cease. This design enabled enhancing spatial perception, while preventing participants from aligning their heads with the sound source while it was active. Such alignment would lead to the consistent usage of the HRTF associated with the 0,0 direction for all presented directions.
Note that Headphone Equalization (HpEq) were not applied due to the dependency of HpEq on the specific HRTFs, which were not available for the HRTFs utilized in the experiment.

The real-time rendering of spatial audio was achieved through direct Head-Related Impulse Response (HRIR) rendering, with all impulse responses for different directions stored and selected for convolution with the signal in real time, based on the head orientation data from the head tracker. The convolution for the binaural signal at each ear is expressed as \(s_{\text{left/right}}(t) = s(t) \ast HRIR_{\text{direction}}(t)\), where \(s(t)\) is the signal and \(HRIR_{\text{direction}}(t)\) is the impulse response corresponding to the tracked direction.
Barycentric interpolation was employed to generate the dynamic listening experience, allowing for smooth transitions between the HRIRs as participants moved their heads. This method effectively provided continuous spatial sound localization cues.
\newline The study involved eight male participants, ranging in age from 25 to 60. All participants had a background in listening tests and spatial hearing. These participants self-reported normal hearing status.

\begin{figure}
    \includegraphics[width=0.50\textwidth,keepaspectratio]{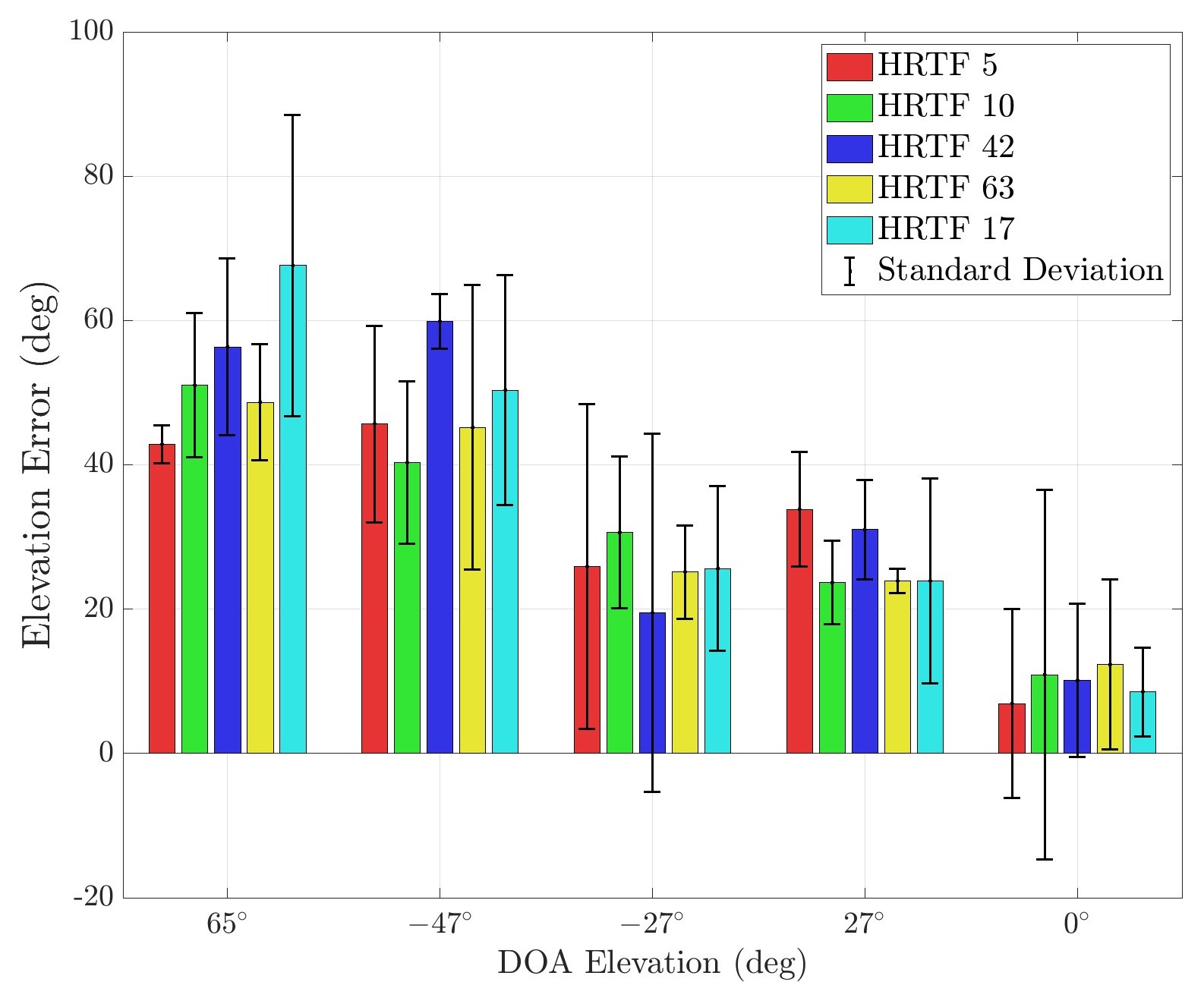}
    \caption{Elevation errors of subject 1}
    \label{fig:subj graph}
\end{figure}

\section{Results}

\subsection{Three methods for HRTF Selection}
The absolute localization errors in azimuth and elevation were analyzed using three distinct methods:
\newline\textbf{Random HRTF}: Randomly selecting one of the five available HRTFs for each of the eight subjects. 
\newline\textbf{Best single HRTF}: Selecting the HRTF that, on average across DOAs and repetitions, yielded the minimal error (in either elevation or azimuth).
\newline\textbf{Best HRTF in each DOA}: Selecting for each subject and for each DOA the HRTF that on average over the repetitions yielded the minimal error. 



In the evaluation process, the absolute azimuth and elevation localization errors for each HRTF and participant in the experiment were initially computed. An illustration of the elevation error results for subject 1 is presented in Figure \ref{fig:subj graph}. This figure displays the elevation error as a function of the source's Direction of Arrival (DOA), with separate representations for each HRTF. It provides insight into the motivation behind our current research, which aims to determine whether the HRTF that results in the smallest localization errors varies across different DOAs. For instance, at DOA elevation 0, the HRTF of subject number 5 yielded the smallest error on average, while at DOA -27 degrees,the HRTF of subject number 42 exhibited the smallest errors. These initial findings motivated the investigation of direction-dependent HRTF selection. 

As previously outlined, three approaches for HRTF selection were performed. For each of the methods, analysis of the elevation and azimuth localization errors was performed.

\begin{figure}
    \centering
     \includegraphics[width=0.5\textwidth]{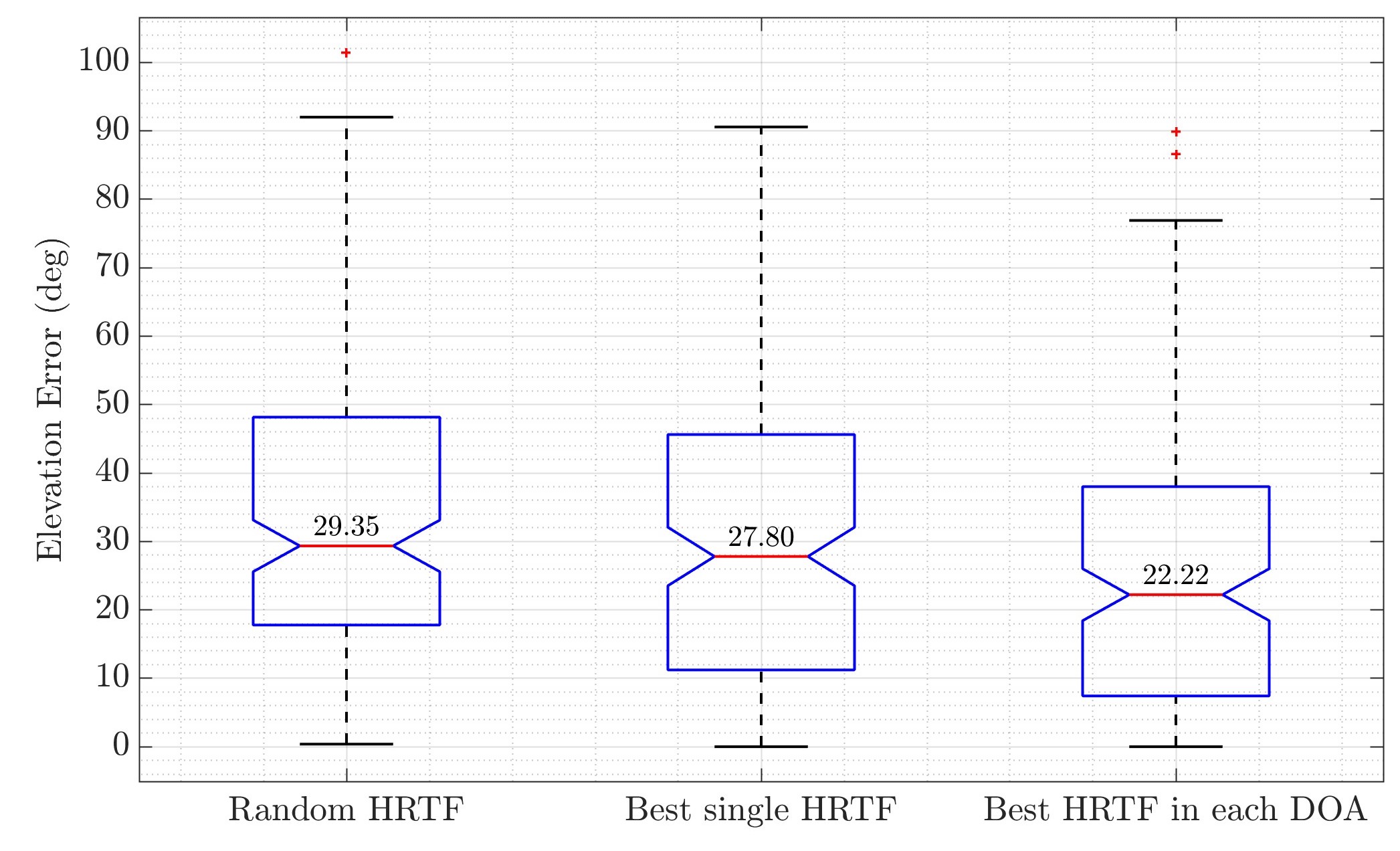}
     \caption{Elevation error in degrees for the three HRTF selection methods}
     \label{fig:boxplot elevation}
 \end{figure}

\begin{figure}
\centering
    \includegraphics[width=0.50\textwidth]{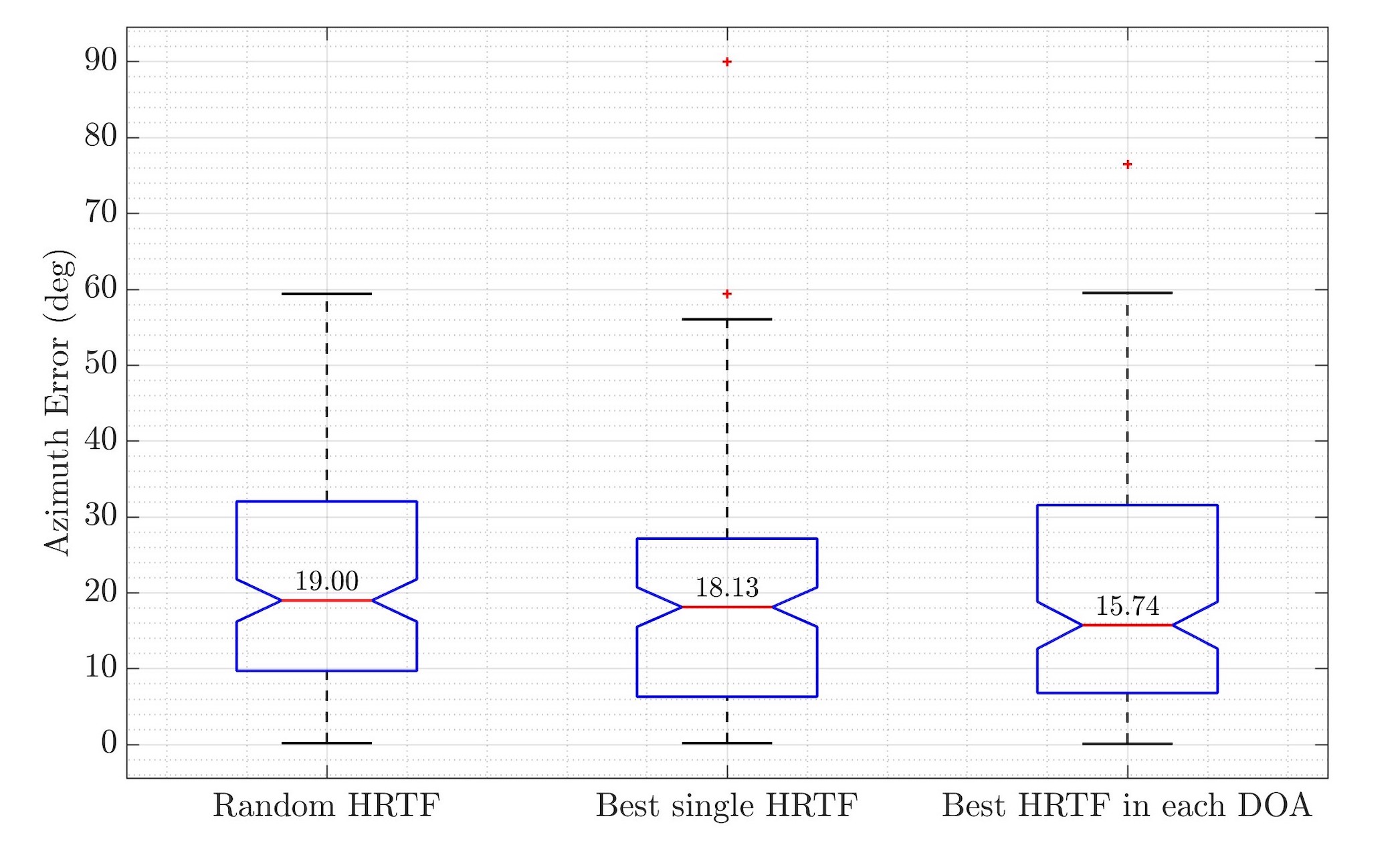}
    \caption{Azimuth error in degrees for the three HRTF selection methods}
    \label{fig:boxplot azimuth}
\end{figure}

The outcomes for the analysis of elevation error are shown in Figure \ref{fig:boxplot elevation}, while the results for azimuth errors, adjusted to project all subject localization evaluations onto the front hemisphere, are displayed in Figure \ref{fig:boxplot azimuth}. It should be noted that azimuth error analysis focuses exclusively on two symmetric azimuth directions, specifically 40 degrees and 320 degrees (left and right).

\subsection{Statistical Analysis}

A Kruskal-Wallis non-parametric test was conducted to compare the efficacy of two of the methods described above: the selection of the best-fitting HRTF on average, versus the approach of individually selecting the best-fit HRTF for each DOA. For elevation errors, A very close to statistically significant difference with a p-value of 0.056 was observed, which was associated with a 5.6-degree difference in the medians  (27.8 degrees for the method of selecting a single HRTF compared to 22.2 degrees for the best-fit in each DOA). Additionally, for azimuth errors, an insignificant  difference with a p-value of 0.12 was noted in conjunction with a 2.4-degree difference in the medians (18.13 degrees for the method of selecting a single HRTF compared to 15.74 degrees for the best-fit in each DOA). Moreover, an insignificant 0.87 -degree difference  in the medians was observed between the method of randomly choosing an HRTF and selecting the averaged best one for the elevation errors, with a p-value of 0.96.

\subsection{Directional Variability of Selected HRTFs}

This chapter analyzes the distribution patterns of HRTFs identified as the best fit for elevation sound localization. Figure \ref{fig:bar_plot_distribution} illustrates the averaged, ordered distribution of these HRTFs, highlighting how frequently each was determined to be the best fit across various directions. For example, the HRTF most commonly identified as providing the best fit was found to be the best for an average of approximately 2.5 out of 5 possible directions. In contrast, in a scenario where one HRTF is the best fit in all directions, the distribution would be expected to feature a single column at the full count of 5. However, it is observed from our data that the distribution neither peaks at a single point nor is uniformly flat.

\section{Discussion and Conclusion }
\label{sec:majhead}

Our research findings strongly support the concept of direction-dependent HRTF selecion, particularly for elevation sound localization. We observed a substantial 5.6-degree difference in the medians of elevation localization errors between employing a direction-dependent HRTF application and using a single, universally optimal HRTF across all directions. This notable difference, underscored by a low p-value of 0.056, suggests a substantial benefit in selecting the most appropriate HRTF for each elevation.

\begin{figure}
\includegraphics[width=0.5\textwidth]{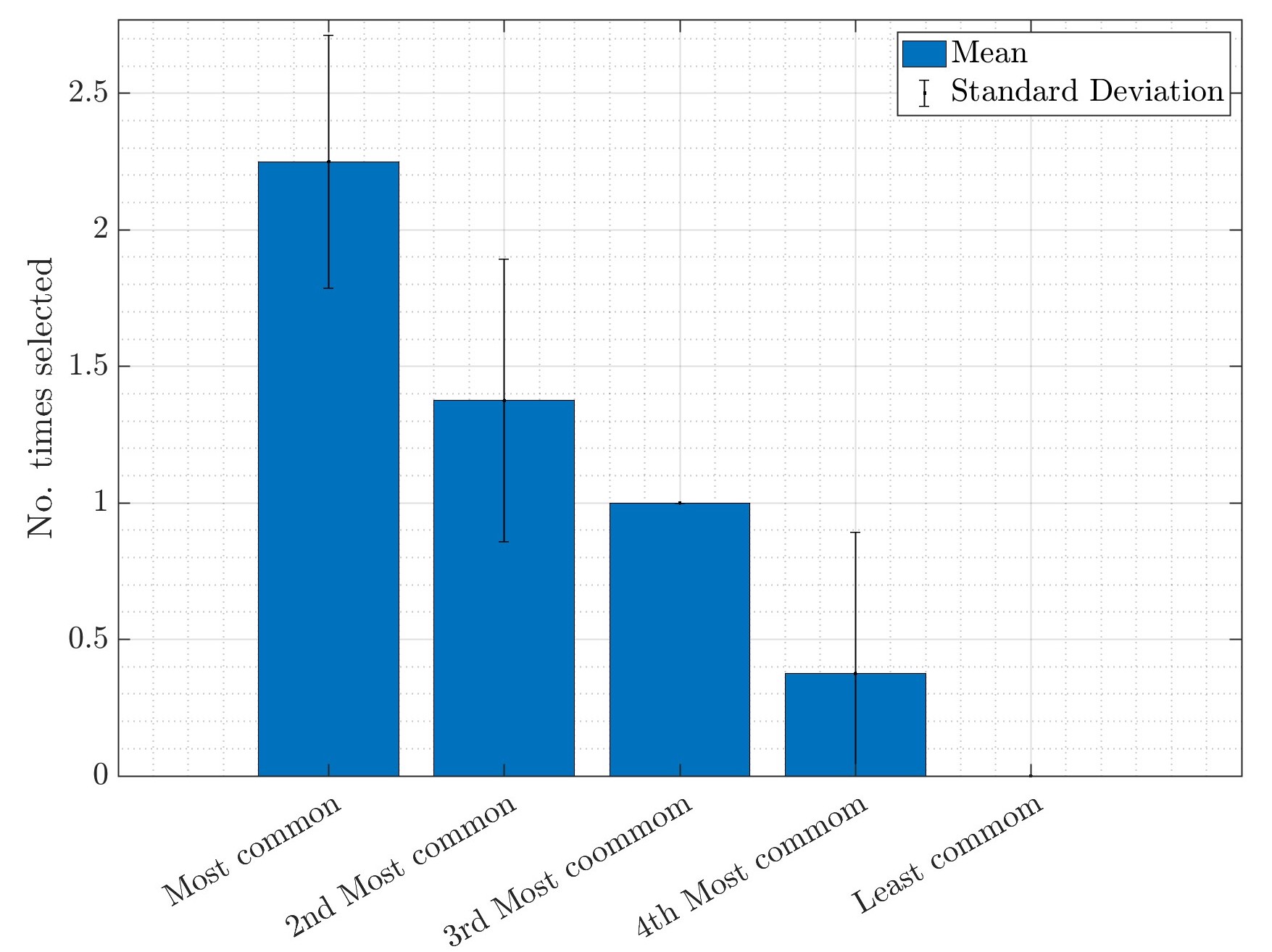}
\caption{Averaged, ordered distribution of HRTFs based on their selection as the optimal fit across different directions. The x-axis represents the rank order of HRTFs by their frequency of being chosen as the best fit, from the most to the least common.}
\label{fig:bar_plot_distribution}
\end{figure}

The distribution analysis of HRTFs' directional optimality further reinforces this conclusion. As depicted in \ref{fig:bar_plot_distribution}, the variability in HRTF optimality across directions suggests no single HRTF consistently outperforms others across all tested elevations. Instead, the most commonly selected HRTF provided the best fit for an average of approximately 2.5 out of 5 directions. This indicates a nuanced preference for certain HRTFs in specific spatial orientations, necessitating the use of multiple HRTFs to achieve optimal localization accuracy. 

Conversely, our study revealed a slightly different scenario when examining azimuth errors. Selecting the optimal HRTF for each azimuth direction had less impact than on elevation errors, with a 2.4-degree difference in the medians and a p-value of 0.12. This outcome was anticipated, given that azimuth localization relies not only on monaural HRTF cues but also on binaural cues, such as the Interaural Time Difference (ITD) and Interaural Level Difference (ILD) \cite{seeber2003subjective}. These binaural cues exhibit less variability among individuals compared to monaural cues, which are heavily influenced by the distinctive morphology of the pinnae and have a greater effect on elevation perception \cite{romigh2014you}. Therefore, while our study varied the HRTFs to assess their impact on localization errors, the inherent characteristics of azimuth localization—primarily governed by these robust binaural cues—meant that changes in HRTF selection method had a smaller effect on azimuth errors.

Additionally, it is noteworthy that azimuth localization errors were relatively low compared to elevation errors. This reduction in errors was achieved by projecting all measurements onto the front hemisphere, which effectively minimized the impact of front-back confusion—a common issue with non-individualized HRTFs \cite{hu2008hrtf}. The smaller error magnitude can also be attributed to the robust binaural cues, which significantly aid azimuth localization.

The comparison of individualized HRTFs with synthesized HRTFs derived by the method studied in this work, is proposed for future research. 
Further exploration of the challenges involved in generating and effectively using 'mixed HRTFs', where for each selected direction HRTF from different individuals may be used, will be integral to understanding the real-world implications of our findings. These challenges may include finding the appropriate method for interpolating all of these HRTFs, the efficiency of the calculations, and the performance of the listening test.

\section{Acknowledgments}
This work was partially supported by Reality Labs @Meta.

\bibliographystyle{IEEEtran.bst}
\bibliography{refs}

\end{document}